\documentclass[%
 reprint,
superscriptaddress,
 amsmath,amssymb,
 aps,
]{revtex4-2}
\usepackage{graphicx}
\usepackage{dcolumn}
\usepackage{bm}
\usepackage{xcolor}
\usepackage{graphicx}     
\usepackage[labelformat=simple]{subcaption}
\usepackage{natbib}
\usepackage{float}
\usepackage{amsmath}
\usepackage{amssymb}

\usepackage{nag}
\usepackage{graphicx}
\usepackage{amsthm}
\usepackage{tabulary}
\usepackage{threeparttable}
\usepackage{qcircuit}
\usepackage{mathtools}
\usepackage{bm}
\usepackage[table,xcdraw]{xcolor}
\usepackage{datetime}
\usepackage{stackrel}
\usepackage{stmaryrd}
\usepackage{xcolor} 
\usepackage{dsfont}

\usepackage[english]{babel}
\makeatletter
\expandafter\def\csname l@en\endcsname{\csname l@english\endcsname}
\expandafter\def\csname dateen\endcsname{\csname dateenglish\endcsname}
\expandafter\def\csname extrasen\endcsname{\csname extrasenglish\endcsname}
\expandafter\def\csname noextrasen\endcsname{\csname noextrasenglish\endcsname}
\makeatother
\usepackage{units}

\usepackage{tikz}
\usetikzlibrary{backgrounds,decorations.pathreplacing}

\usepackage{chngcntr}

\usepackage[normalem]{ulem}
\usepackage{slashed}
\usepackage{soul}

\usepackage{xcolor}
\usepackage[
    colorlinks=true,
    urlcolor=blue,
    linkcolor=blue,
    citecolor=blue,
    filecolor=blue,
]{hyperref}
\usepackage{amssymb}
\usepackage{amsmath}
\usepackage{amsthm}

\usepackage[OT2,T1]{fontenc}
\DeclareSymbolFont{cyrletters}{OT2}{wncyr}{m}{n}
\DeclareMathSymbol{\Sha}{\mathalpha}{cyrletters}{"58}

\usepackage{makeidx}
\makeindex

\usepackage{soul}
\usepackage{xargs}
\newcommandx{\cmnote}[2][1=]{\linespread{1.0}\todo[linecolor=red,backgroundcolor=red!25,bordercolor=red,#1]{#2}}

\let\underline\ul

\allowdisplaybreaks[4]

\DeclareMathOperator{\Tr}{Tr}

\DeclareMathOperator{\diag}{diag}

\let\originalleft\left
\let\originalright\right
\renewcommand{\left}{\mathopen{}\mathclose\bgroup\originalleft}
\renewcommand{\right}{\aftergroup\egroup\originalright}

 \index{} \index{} \index{} \index{} \index{} \index{} \index{} \index{}%
\makeatletter

\newcommand{\ringplus}{\mathbin{\text{\@ringplus}}}

\newcommand{\@ringplus}{%
  \ooalign{\hidewidth\raise1.3ex\hbox{\tiny$\circ$}\hidewidth\cr$\m@th+$\cr}%
}

\newcommand{\ringminus}{\mathbin{\text{\@ringminus}}}

\newcommand{\@ringminus}{%
  \ooalign{\hidewidth\raise0.9ex\hbox{\tiny$\circ$}\hidewidth\cr$\m@th-$\cr}%
}
\makeatother
 \index{} \index{} \index{} \index{} \index{} \index{} \index{} \index{}%

\DeclareFontFamily{U}{wncy}{}
\DeclareFontShape{U}{wncy}{m}{n}{<->wncyr10}{}
\DeclareSymbolFont{mcy}{U}{wncy}{m}{n}
\DeclareMathSymbol{\Sh}{\mathord}{mcy}{"58}

\newcommand{\op}[1]{\hat{#1}}

\renewcommand{\vec}[1]{\bm{#1}}

 \index{} \index{} \index{} \index{} \index{} \index{} \index{} \index{}%
 \index{} \index{} \index{} \index{} \index{} \index{} \index{} \index{}%

 \index{} \index{} \index{} \index{} \index{} \index{} \index{} \index{}%
 \index{} \index{} \index{} \index{} \index{} \index{} \index{} \index{}%
\usepackage{graphicx}

\usepackage{graphicx}
\usepackage{multirow}
\usepackage{hhline}

\xyoption{color}
\xyoption{line}

\newcommandx*\bsbal[3][1=black, 3=->]{\ar @[#1]@{#3} [#2,0] \qw}

\newcommandx*\varbs[4][1=black, 3=\theta, 4=->]{\ar @[#1]@{#4}^{#3} [#2,0] \qw}
\newcommandx*\varbss[4][1=black, 3=\frac{\pi}{4}, 4=->]{\ar @[#1]@{#4}^{#3} [#2,0] \qw}
\newcommandx*\varbsleft[5][1=black, 3=\theta', 4=->]{\ar @[#1]@{#4}^{#3}_{#5} [#2,0] \qw}

\providecommandx*\ctrlg[2]{\control \ar @{-}^{#1} [#2,0] \qw}
\providecommandx*\ctrlog[2]{\controlo \ar @{-}^{#1} [#2,0] \qw}

\makeatletter
 \newcommand{\xmapsfrom}[2][]{%
    \ext@arrow3095\leftarrowfill@{#1}{#2}\mapsfromchar
}
\makeatother

\usepackage{mathtools}

\DeclarePairedDelimiter{\bra}{\langle}{\rvert}%
\DeclarePairedDelimiter{\ket}{\lvert}{\rangle}%
\DeclarePairedDelimiterX\innerp[2]{\langle}{\rangle}{#1\delimsize\vert\mathopen{}#2}%
\DeclarePairedDelimiterX\braketOP[3]{\langle}{\rangle}{#1\,\delimsize\vert\,\mathopen{}#2\,\delimsize\vert\,\mathopen{}#3}%
\DeclarePairedDelimiterX\ketbra[2]{\lvert}{\rvert}{#1\delimsize\rangle\!\delimsize\langle#2}%
\DeclarePairedDelimiterX\outerp[2]{\lvert}{\rvert}{#1\delimsize\rangle\!\delimsize\langle#2}%
\DeclarePairedDelimiterX\projector[1]{\lvert}{\rvert}{#1\delimsize\rangle\!\delimsize\langle#1}%
\usepackage{leftindex}
\DeclareMathOperator*{\argmax}{arg\,max}
\begin{document}

\preprint{APS/123-QED}

\title{All-optical quantum memory using bosonic quantum error correction codes}
\author{Kaustav Chatterjee}
\thanks{These authors contributed equally to this work.}
\email{kauch@dtu.dk}
\affiliation{Center for Macroscopic Quantum States (bigQ), Department of Physics,
Technical University of Denmark, Building 307, Fysikvej, 2800 Kgs. Lyngby, Denmark
}
\author{Niklas Budinger}
\thanks{These authors contributed equally to this work.}
\email{nbudinge@t-online.de}
\affiliation{Center for Macroscopic Quantum States (bigQ), Department of Physics,
Technical University of Denmark, Building 307, Fysikvej, 2800 Kgs. Lyngby, Denmark
}
\affiliation{Johannes-Gutenberg University of Mainz, Institute of Physics, Staudingerweg 7, 55128 Mainz, Germany}
\author{Kian Latifi Yaghin}
\author{Lucas Borg Clausen}
\author{Ulrik Lund Andersen}
\email{ulrik.andersen@fysik.dtu.dk}
\affiliation{Center for Macroscopic Quantum States (bigQ), Department of Physics,
Technical University of Denmark, Building 307, Fysikvej, 2800 Kgs. Lyngby, Denmark
}

\date{\today}

\begin{abstract}

Reliable quantum memory is essential for scalable quantum networks and fault-tolerant photonic quantum computing. We present a quantitative analysis of an all-optical quantum memory architecture in which a Gottesman–Kitaev–Preskill (GKP) encoded qubit is stored in a fibre loop and periodically stabilized using teleportation-based error correction. By modelling fibre propagation as a pure-loss channel and representing each correction round as an effective logical map acting on the Bloch vector, we obtain a compact description of the full multi-round memory channel. We show that syndrome decoder optimization plays a crucial role in the experimentally relevant finite-squeezing regime. The optimal decoder deviates from standard square-grid GKP decoder in both tile-size and tile-shape, leading to significant improved logical performance. Using this optimized decoding strategy, we identify a squeezing-dependent optimal spacing between correction nodes that maximizes the memory lifetime. Remarkably, this optimal segment length is largely independent of the desired storage time, providing a simple and practical design rule for fibre-loop quantum memory. We further find a squeezing threshold of approximately $6.7\,\mathrm{dB}$ below which intermediate error correction becomes counterproductive, while above threshold the achievable storage time increases approximately exponentially with squeezing. For example, at 17 dB squeezing, storage times exceeding $400 \;\mathrm{ms}$ can be achieved with logical infidelity below 1\%. These results establish clear performance benchmarks and reveal the fundamental trade-off between photon loss, squeezing, and correction frequency in continuous-variable architectures. Our findings provide actionable design principles for near-term photonic quantum memory and clarify the path toward scalable all-optical fault-tolerant quantum storage.
\end{abstract}

\maketitle


\section{\label{sec:level1}Introduction}
Quantum memory serves as a central ingredient of scalable quantum technologies. It is required both for synchronizing operations in quantum networks and for buffering information in fault-tolerant quantum computing architectures. In photonic platforms, an especially appealing possibility is an \emph{all-optical} memory, in which quantum information remains encoded and propagating through delay-lines and can therefore interface naturally with existing fibre infrastructure \cite{lvovsky2009optical,Simon2010, Knill2001}. The central obstacle, however, is that optical storage is unavoidably limited by photon loss, which continuously degrades the encoded state during propagation.

Bosonic quantum error correction provides a natural route to mitigating this degradation by encoding a qubit into a single harmonic-oscillator mode rather than a two-level system \cite{Albert2018,Grismo2020,Terhal2020,Cai2021}. Among bosonic encodings, the Gottesman--Kitaev--Preskill (GKP) code is particularly attractive for optical implementations because it converts small phase-space displacement errors into discrete logical errors at the encoded-qubit level \cite{GKP,Grismo2021}. Since a broad class of bosonic noise processes can be represented or well approximated in terms of such displacements, GKP encoding offers a powerful and conceptually clean framework for protecting quantum information in photonic systems \cite{Albert2018,Terhal2020}.

At the same time, finite-resource GKP error correction is itself noisy. In realistic architectures, the ancillary GKP resource states used for teleportation-based correction have finite energy, so each correction step suppresses accumulated propagation errors while also injecting fresh noise into the logical state \cite{Knill2006,Menicucci,Hastrup}. This leads to a fundamental and practically relevant design question: when does repeated error correction improve memory performance, and when does it instead become counterproductive? Answering this question is essential for assessing whether all-optical GKP memories can provide a practical advantage in realistic fibre-based settings.

In this work, we study an all-optical quantum-memory architecture in which a GKP-encoded optical mode is stored in a fibre loop and periodically stabilized using teleportation-based GKP error correction, while corrective Pauli operations are tracked classically and applied only once at readout. We model fibre propagation as a pure-loss channel and represent each correction round by a syndrome-conditioned effective logical map acting on the encoded qubit. This yields a compact description of the full multi-round memory channel and allows us to optimize the decoder directly for the finite-squeezing, finite-loss regime relevant to experiments, rather than fixing it to the ideal nearest-lattice GKP rule \cite{Jafarzadeh_2025}. In this sense, decoding is not merely a classical post-processing step, but an integral part of the physical memory design.

Our analysis reveals three main results. First, decoder optimization leads to a systematic and quantitatively significant improvement in logical performance in the experimentally relevant finite-squeezing regime. Second, for a given squeezing resource, there exists an optimal spacing between correction nodes that maximizes memory performance; remarkably, this spacing is largely independent of the total target storage time, thereby providing a simple and practical design rule for fibre-loop memories. Third, the protocol exhibits a clear squeezing threshold near \(6.7\,\mathrm{dB}\): below this threshold, intermediate error correction is detrimental, whereas above it the achievable storage time increases rapidly with squeezing. This threshold marks the transition between regimes where correction is beneficial and where it becomes counterproductive. In our optimized setting, storage times exceeding \(400\,\mathrm{ms}\) with logical infidelity below \(1\%\) are achievable at \(17\,\mathrm{dB}\) squeezing.

\begin{figure}[t]
    \centering
    
    
    \includegraphics[width=\linewidth]{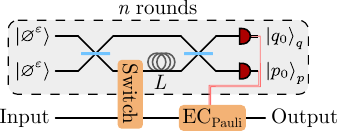}
    \caption{Experimental setup of the proposed quantum memory. A GKP-encoded optical mode is switched into a fibre loop of length $L$ where it is subjected to photon loss. Next, teleportation-based GKP error correction -- consisting of two approximate qunaught states $\left|\varnothing^\varepsilon\right>$, two beam-splitters (blue) and two homodyne detectors (red) -- projects the state back into the approximate GKP code space and corrects loss-induced errors. After passing the fibre loop $n$ times, the mode is switched out and a logical Pauli correction based on the accumulated homodyne outcomes is applied.}
    \label{fig:conceptart}
\end{figure}
These results place all-optical GKP memories on a more quantitative footing. Rather than asking only whether bosonic encoding can protect optical information in principle, they identify the operating regime in which repeated correction becomes beneficial and extract experimentally relevant design principles for how such a memory should be engineered. More broadly, they clarify how finite-resource bosonic error correction should be deployed in photonic architectures where the act of correcting noise is itself noisy. In this way, the work establishes a direct link between physical resources (such as squeezing) and achievable memory performance. They also point toward a natural next step: concatenating the GKP layer with a higher-level qubit code once sufficiently good GKP resources and photonic hardware become available. The rest of the paper is organized as follows. In Sec.\,\ref{sec:level2} we review the GKP code, finite-energy codewords, and teleportation-based error correction under a pure-loss model. In Sec.\,\ref{sec:results} we present the loop-memory protocol and benchmarking metric, introduce the fidelity-optimized decoder and its phase-space structure, and then analyse the resulting trade-offs between squeezing, loss, correction spacing, and storage time. Finally we conclude with a discussion and outlook in Sec. \ref{sec4}.

\section{\label{sec:level2}Preliminaries} 

\subsection{The GKP code}
We start with a brief review of the GKP code \cite{GKP, Menicucci}. Specifically, we consider the square GKP code which encodes a qubit into a single-mode continuous-variable (CV) Hilbert space $\mathcal{H}$, described by position ($\hat{q}$) and momentum ($\hat{p}$) operators satisfying $[\hat{q},\hat{p}]=i$. The ideal GKP codespace is the simultaneous $(+1)$-eigenspace of the two commuting stabilizers
\begin{equation}
\hat S_1 = e^{-2 i \sqrt{\pi}\,\hat p}, 
\qquad
\hat S_2 = e^{ 2 i \sqrt{\pi}\,\hat q}
\end{equation}
and is spanned by the ideal codewords
\begin{equation}
|0\rangle = \sum_{s\in\mathbb Z} |(2s)\sqrt{\pi}\rangle_q,
\qquad
|1\rangle = \sum_{s\in\mathbb Z} |(2s+1)\sqrt{\pi}\rangle_q,
\end{equation}
where $|q_0\rangle_q$ denotes the $\hat q$-eigenstate with eigenvalue $q_0$.
These ideal codewords are unphysical because they possess infinite energy. Nevertheless, they provide a useful starting point for constructing finite-energy approximations of the code states. physically valid states by applying the damping operator
\begin{equation}
\hat N(\varepsilon) = e^{-\varepsilon \hat n},
\qquad
\hat n=\tfrac12(\hat q^2+\hat p^2-1),
\end{equation}
where the parameter $\varepsilon$ is related to the usual dB squeezing value via $-10\log_{10}(\tanh \varepsilon)$ \cite{matsuura_equivalence_2020}. The approximate GKP codewords are then typically given by
\begin{equation}
|\bar 0^\varepsilon\rangle \propto \hat N(\varepsilon)\,|0\rangle,
\qquad
|\bar 1^\varepsilon\rangle \propto \hat N(\varepsilon)\,|1\rangle
\end{equation}
along with the proper normalisation $\langle \bar 0^\varepsilon|\bar 0^\varepsilon\rangle=\langle \bar 1^\varepsilon|\bar 1^\varepsilon\rangle=1$. However, these two physical codewords do not form an orthonormal basis, as they are not orthogonal, 
\begin{equation}
\delta \equiv \langle \bar 0^\varepsilon|\bar 1^\varepsilon\rangle \neq 0.
\end{equation}
For analytical convenience it is therefore useful to introduce an orthonormal basis constructed from symmetric and antisymmetric combinations of the approximate codewords,
\begin{equation}
|+^\varepsilon\rangle = \frac{|\bar 0^\varepsilon\rangle+|\bar 1^\varepsilon\rangle}{\sqrt{2(1+\delta)}},
\qquad
|-^\varepsilon\rangle = \frac{|\bar 0^\varepsilon\rangle-|\bar 1^\varepsilon\rangle}{\sqrt{2(1-\delta)}},
\end{equation}
which form an orthonormal basis by construction. 
In the high-squeezing regime $\varepsilon\ll 1$, the overlap $\delta$ becomes negligible and the two bases effectively coincide.
For low squeezing, on the other hand, this choice provides a convenient orthonormal representation of the approximate GKP codespace, allowing us to treat the space of approximate GKP states as an effective logical qubit without introducing additional approximations.
Consequently, any mixed approximate GKP state represented by the density matrix $\hat \rho^\varepsilon$ can be concisely represented by the Bloch vector $\vec a=(1, a_1,a_2,a_3)^T$ as
\begin{equation}
\hat \rho^\varepsilon = \frac{1}{2}\vec a \cdot\hat{\vec \sigma}^\varepsilon
\end{equation}
with the vector of Pauli matrices $\hat{\vec\sigma}^\varepsilon$, defined in the orthonormal approximate-GKP basis, where $\hat \sigma_0^\varepsilon$ is the identity.

\subsection{Loss and GKP error correction}
\label{sec:loss_gkp_ec}

\begin{figure}[t]
    \centering
    \includegraphics[width=1\linewidth]{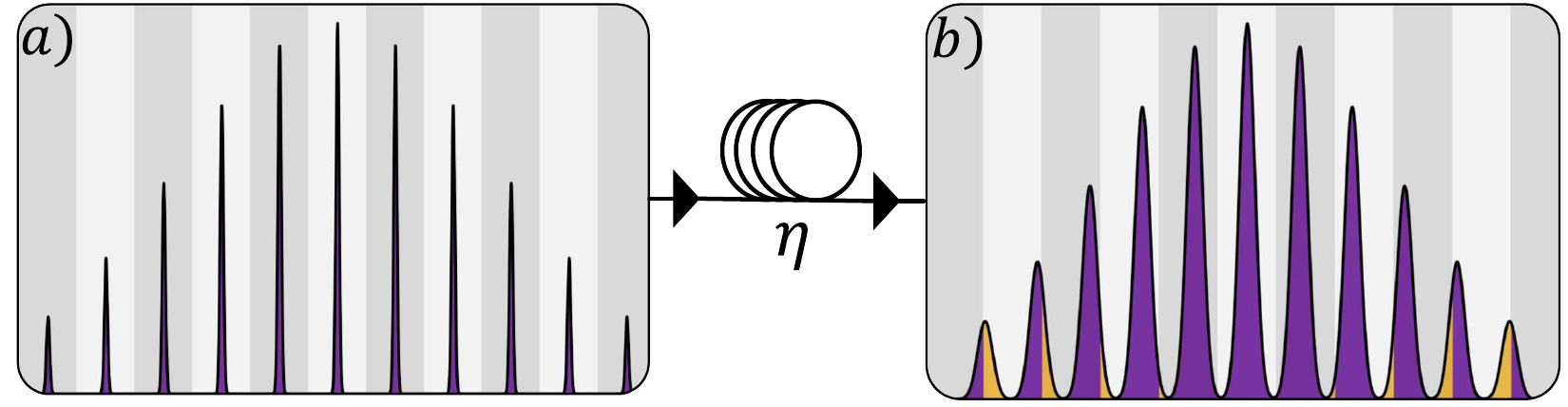}
    \caption{Illustration of the effect of a pure-loss channel on a GKP state. The input a) is an approximate GKP $\ket{+^\varepsilon}$ state.
    Grey and white stripes indicate logical bins. b) After transmission through the pure-loss
    channel, the peaks move closer to the origin and simultaneously broaden. Peaks spilling into neighbouring cells and causing logical errors are highlighted in orange.}
    \label{Peak_broading_and_shifting}
\end{figure}

Our quantum-memory protocol consists mainly of two alternating steps: (i) storing a GKP-encoded qubit in an optical fibre and (ii) periodically performing
teleportation-based GKP error correction.
The optical fibre and the accompanying photon loss are modelled as a pure-loss channel $\mathcal L_\eta$ with Stinespring dilation
\begin{equation}
\mathcal{L}_\eta(\hat \rho_S)=\mathrm{tr}_E\!\Big(\op{B}_\eta(\hat \rho_S\otimes \ket{\mathrm{vac}}\!\bra{\mathrm{vac}}_E)\op{B}_\eta^\dagger\Big),
\end{equation}
where $\op{B}_\eta$ denotes a beam-splitter unitary coupling the system mode ($S$) to an environmental vacuum mode ($E$),
\begin{equation}
\begin{split} \label{cir:beamsplitter}
	\raisebox{-1.2em}{$\op{B}_{\eta}  =
 e^{ - i\theta ( \op{q}_S \op{p}_E - \op{p}_S \op{q}_E )} \, = \, $}
         \Qcircuit @C=1.25em @R=2.5em @! 
         {
          &  \varbs{1} & \rstick{S}  \qw \\
          & \qw       & \rstick{E} \qw
  		  } 
\end{split}\quad\quad,
\end{equation}
where $\cos^2(\theta)=\eta$ defines the transmissivity of the beam-splitter.
The effect of loss on an approximate GKP state is twofold. First, the variance of the Gaussian peaks increases due to the added vacuum noise. Second, the individual peaks are displaced towards the origin relative to their initial position in phase space. Both effects lead to logical errors by pushing probability distributions into wrong quadrature bins as illustrated in Fig.~\ref{Peak_broading_and_shifting}. However, the relative importance of these two effects depends strongly on the squeezing level of the GKP state. For low squeezing, the increased peak variance is the dominant source of logical errors, whereas in the high-squeezing regime the dominating mechanism is the inward displacement of peaks located far from the origin. In \cite{Hastrup} this was shown to lead to an optimal squeezing value $\varepsilon$ for a given amount of loss $\eta$. In principle, amplification could be used to partially compensate for the inward peak displacement, but it has been shown to generally perform worse than the unamplified case~\cite{Hastrup}.

To manage the CV noise introduced by photon loss, we use teleportation-based GKP error correction \cite{Menicucci}.
The key resource of this protocol is the GKP qunaught state, defined as
\begin{align}
    \ket{\varnothing}=\sum_{s\in \mathbb{Z}}\ket{s\sqrt{2\pi}}_q.
\end{align}
When two such states are mixed on a balanced beam-splitter, they create a GKP Bell pair $\hat B_{\frac{\pi}{4}}\ket{\varnothing\varnothing}\propto\ket{00}+\ket{11}$. Together with a Bell measurement consisting of another balanced beam-splitter and two homodyne measurements, this resource enables the teleportation-based error-correction circuit:
\begin{equation} 
\begin{split} \label{Cir:teleport}
    \Qcircuit @C=2.0em @R=2.5em  
    {
    &&&\lstick{\text{(in)}}  &\qw & \varbss{1} &  \qw&\ket{q_0}_q \\
    &&&\lstick{\ket{\varnothing^\varepsilon}}  & \varbss{1} &  \qw & \qw&\ket{p_0}_p\\
    &&&\lstick{\ket{\varnothing^\varepsilon}}  & \qw &  \qw & \qw & (\text{out})
    }
\end{split}
\qquad\qquad
\end{equation}
Here, the approximate qunaught states are given by $\ket{\varnothing^\varepsilon}\propto\hat N(\varepsilon)\ket{\varnothing}$ with $\langle \varnothing^\varepsilon|\varnothing^\varepsilon\rangle=1$.
The output state of this circuit can be expressed in terms of logical coefficients, see App.~\ref{sec:app}. Operationally, the teleportation-based GKP error correction projects the lossy states back into the logical GKP codespace and converts CV noise into discrete logical Pauli errors. Hence, the application of loss followed by the teleportation can be fully described by an effective qubit operation given by the matrix $T^{\varepsilon, \eta}(q_0,p_0)$ acting on a Bloch vector as
\begin{align}
    \vec{a}_\text{out}(q_0,p_0)=T^{\varepsilon, \eta}(q_0,p_0) \vec{a}_\text{in}.
\end{align}
For a derivation of $T^{\varepsilon, \eta}(q_0,p_0)$ see App.~\ref{sec:app}. Importantly, the induced operation is not necessarily a completely positive trace-preserving (CPTP) map within the logical qubit space, as the projection onto the approximate GKP codespace is not completely positive \cite{Jafarzadeh_2025}.
After the teleportation, the induced logical error must still  be corrected in order to protect the stored quantum information.
To this end, we consider a Pauli correction based on the homodyne outcomes $q_0$ and $p_0$ given by the decoder
\begin{align}
    D^{\varepsilon, \eta}:\mathbb{R}^2\rightarrow \left\{I, X, Y, Z\right\}.
\end{align}
The matrices corresponding to the respective Pauli operators (written as Pauli transfer matrices acting directly on the Bloch vectors) are given by $I=\diag(1,1,1,1)$, $X=\diag(1,1,-1,-1)$, $Y=\diag(1,-1,1,-1)$ and $Z=\diag(1,-1,-1,1)$.
Together, one round of photon loss, GKP error correction and logical error correction is described by the effective transfer matrix
\begin{align}\label{eq:singleround}
    C^{\varepsilon, \eta}\equiv\int_{\mathbb{R}} dq_0\int_{\mathbb{R}} dp_0\ D^{\varepsilon, \eta}(q_0,p_0)T^{\varepsilon, \eta}(q_0,p_0).
\end{align}

\section{Results}
\label{sec:results}

\subsection{Proposed design}

We consider an all-optical memory built from a standard single-mode telecom fibre as illustrated in Fig.~\ref{fig:conceptart}. A travelling optical mode carrying a GKP-encoded qubit is first switched into the fibre loop of length $L$, and subsequently passes through the GKP error correction node of Eq.~\eqref{Cir:teleport}, consisting of two approximate GKP qunaught states with squeezing $\varepsilon$, two balanced beam-splitters and two homodyne detections.
Multiple rounds of error correction can be performed in place corresponding to the creation of a temporally encoded one-dimensional cluster state \cite{PRXQuantum.2.030325}.
After $n$ rounds of alternating fibre loop and GKP error correction, the mode is switched out and a logical Pauli correction is applied. With a speed of light in fibre $c_\text{fib}=2.07\times 10^8 \mathrm{m/s}$ ($n_\text{fib}= 1.45 $) the storage time is given by
\begin{align}\label{eq:storagetime}
    t_s=\frac{n\cdot L}{c_\text{fib}},
\end{align}
while the amount of loss per fibre segment is described by the transmissivity
\begin{align}
    \eta(L) = \exp\left(-\frac{L}{L_{\mathrm{att}}}\right)
\end{align}
with the attenuation length $L_{\mathrm{att}}=22\,\mathrm{km}$ \cite{häussler2024quantumrepeatersbasedstationary}.
Noise due to switching as well as imperfections in beam-splitters and homodyne measurements, is neglected in the present analysis, as our focus lies on the fundamental interplay between errors originating from finite GKP squeezing $\varepsilon$ and photon loss $\eta(L)$ in the fiber. On the other hand, additional coupling loss $\eta_\text{cpl}$ can readily be incorporated using the multiplicativity of loss, giving an effective transmissivity  $\eta_{\text{eff}}(L)=\eta_{\text{cpl}}\cdot\eta(L)$.
Pauli corrections are assumed to be tracked and implemented after the state is switched out. This can either be done physically or applied in post-processing through Pauli-frame tracking, and is considered to be ideal.


\subsection{Figure of merit}

\begin{figure}[t]
    \centering
    \includegraphics[width=1\linewidth]{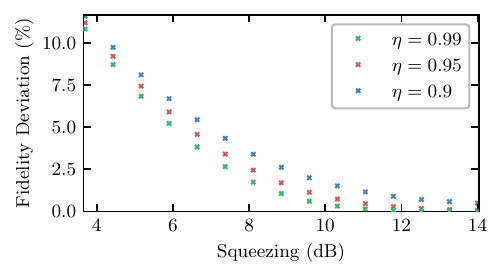}
    \caption{Relative deviation of the worst-case fidelity from the average fidelity for different amounts of photon loss $\eta$ and squeezing.}
    \label{fig:Best_vs_Worst}
\end{figure}

After $n$ rounds of our protocol followed by a Pauli correction, we obtain an effective transformation of the encoded GKP subspace designed to preserve the initial logical information. To benchmark memory performance we compare this effective transformation to the qubit identity channel using the average channel fidelity
\begin{equation}
    F_\text{avg}(\mathcal{E})=\int_{\mathcal{H}_2} d\psi \bra\psi\mathcal{E}(\ket\psi\bra\psi)\ket\psi,
    \label{eq:Favg}
\end{equation}
which measures how well arbitrary logical input states are preserved under a given channel \cite{NielsenChuang2010}. For a qubit channel, this can be reduced to the expression \cite{BOWDREY2002258}
\begin{equation}
    F_\text{avg}(\mathcal{E})=\frac{1}{6}\sum_{i\in\mathcal{B}}\bra{i}\mathcal{E}(\ket i\bra i)\ket{i}
\end{equation}
where the set $\mathcal{B}$ contains the eigenstates of the three Pauli matrices $\hat\sigma^\varepsilon_i$. In terms of Bloch vectors this corresponds to the set $\mathcal B=\{(1,\pm 1,0,0),(1,0,\pm 1,0),(1,0,0,\pm 1)\}$ and we have
\begin{equation}\label{eq:avgBloch}
    F_\text{avg}(\mathcal E)=\frac{1}{2}+\frac{1}{12}\sum_{\vec a_i\in\mathcal{B}}\vec a_{i}\cdot \mathcal{E}(\vec a_{i}).
    \end{equation}
As the effective error-correction process generally introduces anisotropic distortions of the Bloch sphere, some logical states may be preserved better than others. This motivates a comparison with the worst-case fidelity
\begin{equation}
    \begin{split}
        F_\text{wc}&= \min_{\ket{\psi}\in\mathcal{H}_2} \bra{\psi}\mathcal{E}(\ket{\psi}\bra{\psi})\ket{\psi}.
    \end{split}
\end{equation}
The relative deviation between the average and worst-case fidelity for a single round of the protocol is shown in Fig.~\ref{fig:Best_vs_Worst} as a function of squeezing and loss. For fixed $\eta$, the spread between $F_{\mathrm{wc}}$ and
$F_{\mathrm{avg}}$ decreases as the squeezing increases, indicating isotropic behaviour of the effective logical channel at high squeezing. Moreover, the absence of pronounced outliers in performance across logical states supports the further use of the average channel fidelity $F_{\mathrm{avg}}$ as a reliable and convenient benchmark for evaluating the memory performance.

\subsection{Optimized decoder}

\begin{figure}[t]
    \centering
    \includegraphics{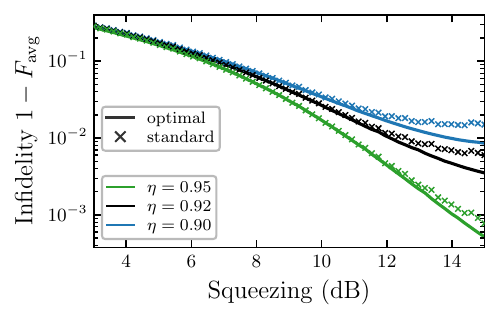}

    \caption{Comparison of the optimal and standard decoder for a single round of teleportation-based GKP error correction for different amounts of loss $\eta$ and squeezing.}
    \label{fig:diffinfid}
\end{figure}

\begin{figure*}[t]
    \centering
    \includegraphics[width=0.329\linewidth]{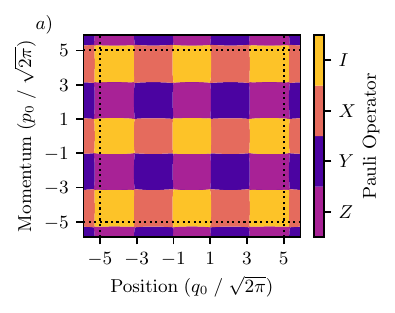}
    \includegraphics[width=0.329\linewidth]{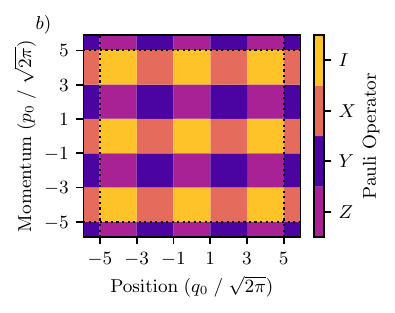}
    \includegraphics[width=0.329\linewidth]{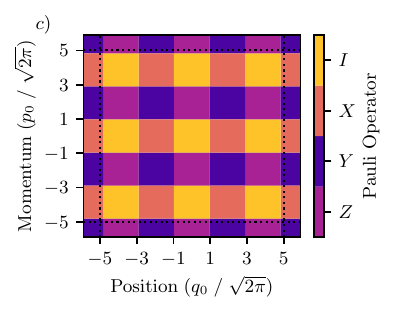}
    \caption{Optimized decoder maps $D_{\mathrm{opt}}^{\varepsilon, \eta}(q_0,p_0)$ for a) $5\mathrm{dB}$, $\eta=1$, b) $10\mathrm{dB}$, $\eta=1$, and c) $10\mathrm{dB}$, $\eta=0.8$. Colours denote the optimal Pauli correction $\{I,X,Y,Z\}$. The dashed grid indicates the standard GKP decoder optimal for high squeezing without loss. In comparison, low squeezing produces wider, distorted squares, while loss contracts regions towards the origin.}
    \label{fig:Checkerboard}
\end{figure*}

The teleportation-based GKP error correction induces a logical error on the stored qubit which depends on the continuous homodyne outcomes $q_0$ and $p_0$. The role of the decoder is to
assign to each outcome the Pauli correction that best restores the logical state. In this work, we choose $D_{\text{opt}}^{\varepsilon, \eta}$ by directly maximizing the average channel fidelity $F_{\text{avg}}$ leading to
\begin{align}
    D_{\text{opt}}^{\varepsilon, \eta}(q_0,p_0)\equiv\argmax_{P\in\{I, X, Y, Z\}}\sum_{\vec a_i\in\mathcal{B}}\vec a_{i}^TPT^{\varepsilon, \eta}(q_0,p_0) \vec a_{i}.
\end{align}
The resulting optimized decoder is visualized in Fig.~\ref{fig:Checkerboard} for different squeezing levels $\varepsilon$ and loss transmissivities $\eta$. In the high-squeezing regime, the optimized decoder approaches the standard $\sqrt{2\pi}$-spaced square-lattice decoder expected for ideal GKP states \cite{Jafarzadeh_2025} (see Fig.~\ref{fig:Checkerboard}b for 10 dB, $\eta=1$). At lower squeezing, intrinsic teleportation noise and finite-energy effects broaden the decision window, distort the characteristic checkerboard pattern and the fidelity-optimizing boundaries become visibly curved (see Fig.~\ref{fig:Checkerboard}a for 5 dB, $\eta=1$). Introducing photon loss further modifies the decoder by contracting the decision regions towards the origin, reflecting the inward displacement of the GKP peaks under loss (see Fig.~\ref{fig:Checkerboard}c for 10 dB, $\eta=0.8$). The quantitative benefit of using the optimized decoder over the standard one for a single teleportation is shown in Fig.~\ref{fig:diffinfid} for different squeezing and loss parameters. The improvement becomes most pronounced when higher squeezing is combined with non-negligible loss, and as the resulting gain accumulates over multiple rounds of the protocol, the optimized decoder is particularly important for long storage times.

\subsection{Quantum memory performance}
\label{subsec:qmem_protocol}

Let us now consider the storage of information over many fibre loops in order to achieve a desired storage time $t_s$. We also assume the squeezing $\varepsilon$ of the GKP qunaught states $\ket{\varnothing^\varepsilon}$ to be fixed, as these states constitute the most experimentally demanding resource in the protocol \cite{larsen_integrated_2025}. However, this still leaves us with the choice of the number of loops $n$. While the total fibre length $L_\text{tot}$ is given by the storage time and Eq.~\eqref{eq:storagetime}, its partitioning into $n$ segments of length $L(n)=\frac{L_\text{tot}}{n}$ has a significant impact on the performance of the memory. On the one hand, performing error correction more frequently introduces the intrinsic teleportation noise at every correction step. On the other hand, performing error correction too infrequently allows loss-induced noise to accumulate during propagation between nodes.
Consequently, we seek to determine the optimal number of segments $n$ that maximizes memory performance. 
In order to accurately calculate these $n$ loops of the memory protocol, we must account for the fact that Pauli corrections are classically tracked and applied only once at readout, as it is experimentally impractical to apply a syndrome informed Pauli operation after each segment individually.

\subsubsection{Keeping track of Pauli corrections}
\label{subsubsec:multi_round_eval}

\begin{figure}
    \centering
    \includegraphics[width=0.65\linewidth]{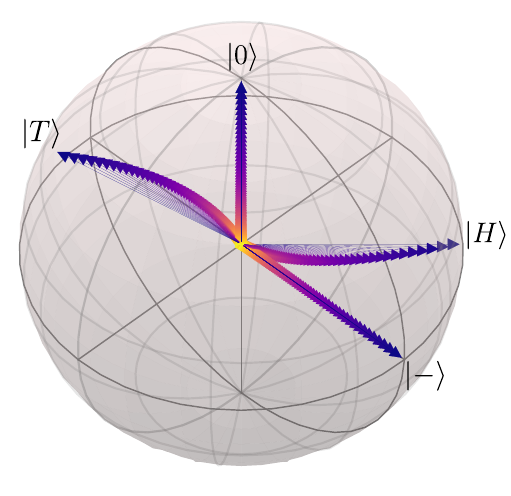}
    \caption{Bloch vector trajectories
    of four distinct GKP qubit states: the $\ket{0^\varepsilon},\ket{-^\varepsilon},\ket{T^\varepsilon}$ and an $\ket{H^\varepsilon}$-like state. The transformation is given by $C^{\varepsilon, \eta}_{(n)}$ for increasing $n$, a fixed squeezing value of $12$dB and $\eta=\eta(L_{\mathrm{opt}}^\varepsilon)$. As the number of rounds increases, the states contract towards the origin while moving closer to the $x$-$z$-plane.}
    \label{fig:bloch}
\end{figure}

Pauli operations do not commute with the teleportation circuit, and therefore the simple product of the single-round matrices in Eq.~\eqref{eq:singleround} $\left(C^{\varepsilon, \eta}\right)^n$, which includes the repeated application of Pauli corrections, does not correctly describe the proposed memory protocol.  Instead, we keep track of the Pauli frame that accumulates during the protocol, representing the necessary but not yet applied Pauli operation. To this end, we define the four matrices
\begin{align}
    P^{\varepsilon,\eta}=\int_{\left(D^{\varepsilon,\eta}_\text{opt}\right)^{-1}(P)} dq_0 dp_0\ T^{\varepsilon,\eta}(q_0,p_0)
     \label{Error matrix E_i}
\end{align}
for $P\in\{I, X, Y, Z\}$ and the pre-image $\left(D^{\varepsilon,\eta}_\text{opt}\right)^{-1}(P)=\left\{(q_0,p_0)\big|D^{\varepsilon,\eta}_\text{opt}(q_0,p_0)=P\right\}$.
Multi-round evolution is then obtained by composing these matrices across rounds according
to the Pauli-frame multiplication rule. This is implemented via the $16\times16$ transfer matrix
\begin{equation}
M^{\varepsilon,\eta} =
\begin{bmatrix}
I^{\varepsilon,\eta} & X^{\varepsilon,\eta} & Z^{\varepsilon,\eta} & Y^{\varepsilon,\eta} \\
X^{\varepsilon,\eta} & I^{\varepsilon,\eta} & Y^{\varepsilon,\eta} & Z^{\varepsilon,\eta} \\
Z^{\varepsilon,\eta} & Y^{\varepsilon,\eta} & I^{\varepsilon,\eta} & X^{\varepsilon,\eta} \\
Y^{\varepsilon,\eta} & Z^{\varepsilon,\eta} & X^{\varepsilon,\eta} & I^{\varepsilon,\eta}
\end{bmatrix}.
\label{eq:M_def}
\end{equation}
Intuitively, the matrix $M^{\varepsilon,\eta}$ describes how Pauli sectors are reassigned when Pauli-frame updates from consecutive rounds are combined.
Hence, the correct transformation matrix of a $n$ round storage followed by a final Pauli correction is
\begin{align}
    C^{\varepsilon, \eta}_{(n)}=
    \begin{bmatrix}
        I&X&Y&Z
    \end{bmatrix}\cdot
    \left(M^{\varepsilon,\eta}\right)^n\cdot
    \begin{bmatrix}
        I&0&0&0
    \end{bmatrix}^T.
\end{align}
Following Eq.~\eqref{eq:avgBloch}, the resulting memory performance is then
\begin{equation}
    F_\text{avg}(\varepsilon, \eta, n)=\frac{1}{2}+\frac{1}{12}\sum_{\vec a_i\in\mathcal{B}}\vec a_{i}^T C^{\varepsilon, \eta}_{(n)}\vec a_{i}.
\end{equation}
On the other hand, the effect of storage over $n$ rounds on a specific state $\vec{a}_\text{in}$ is simply obtained as
\begin{align}
    \vec{a}_\text{out}^{(n)}=C^{\varepsilon, \eta}_{(n)} \vec{a}_\text{in}.
\end{align}
Fig.~\ref{fig:bloch} provides a geometric illustration of the application of $C^{\varepsilon, \eta}_{(n)}$ on a set of Bloch vectors as $n$ increases. Repeated rounds progressively contract the Bloch vectors towards the origin,
indicating an increasingly depolarizing effect. In addition, the trajectories of the $\ket{T}$ and $\ket{H}$-like magic states reveal an asymmetric drift towards the $x$-$z$-plane, demonstrating that the effective logical channel introduces distortions beyond simple depolarization. Accurately tracking these distortions is essential for correctly describing the multi-round memory dynamics, as they cannot be captured by a simple power of an average qubit channel. With this Pauli-frame tracking formalism in place, it becomes possible to evaluate the performance of arbitrary multi-segment memory configurations in a computationally efficient manner.

\subsubsection{Optimal segment length}

\begin{figure}
    \centering
    \includegraphics{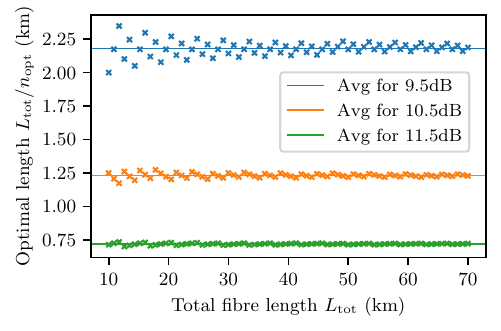}
    \caption{Optimal segment length $L_{\mathrm{tot}}/n_{\mathrm{opt}}$ as a function of total fibre length $L_{\mathrm{tot}}$ for several squeezing values. The horizontal lines indicate the averaged optimal segment length $L_{\mathrm{opt}}^\varepsilon$.}
    \label{fig:LoverN}
\end{figure}

\begin{figure}
    \centering    \includegraphics[width=1\linewidth]{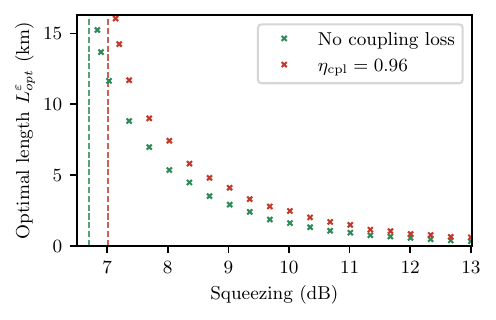}
    \caption{Optimal segment length $L^\varepsilon_{\mathrm{opt}}$ as a function of squeezing in case of no coupling loss (green) and a coupling loss of $\eta_\text{cpl}=0.96$ (red). As squeezing decreases towards a threshold of 6.7dB (7.0dB) indicated by the dashed line, the optimal length $L^\varepsilon_{\mathrm{opt}}$ diverges, highlighting a region of low squeezing where error correction is not beneficial. Coupling losses increase both the optimal segment lengths as well as the threshold.}
    \label{fig:Lopt_vs_squeezing}
\end{figure}

Given a squeezing $\varepsilon$ and a total fibre length $L_{\mathrm{tot}}$, we can now compare memory performance for different choices of segment length $L=\frac{L_\text{tot}}{n}$. In Fig.~\ref{fig:LoverN}, the optimal segment lengths are shown for different squeezing and increasing $L_\mathrm{tot}$. Notably, 
for a given squeezing level, the optimal segment length $L=\frac{L_\mathrm{tot}}{n_{\mathrm{opt}}}$ remains approximately constant (up to finite precision arising from the integer constraint on $n$). This indicates that the optimal segment length is largely independent of the total fiber length and hence of the overall storage time.
Consequently, the optimal fibre partition can be characterized by a squeezing-dependent optimal segment length
$L_{\mathrm{opt}}^\varepsilon$. We determine $L_{\mathrm{opt}}^\varepsilon$ by averaging the numerically obtained optimal segment lengths over a range of total fiber lengths $L_\text{tot}$ as seen in Fig.~\ref{fig:LoverN}. The resulting dependence of the optimal segment length on squeezing is plotted in Fig.~\ref{fig:Lopt_vs_squeezing}. We observe the emergence of a squeezing threshold of approximately $6.7$dB, below which the optimal segment length diverges and the application of error correction becomes counterproductive. In this regime, the noise introduced by teleportation-based GKP error correction outweighs the benefits of correcting the accumulated loss errors. This behavior reflects the fundamental competition between two error mechanisms: loss-induced noise accumulated during fibre propagation and the intrinsic teleportation noise introduced at each error-correction node. Above the threshold, periodic error correction suppresses the growth of loss errors, while below the threshold the added correction noise dominates the dynamics.
Note that the inclusion of coupling losses increases both the optimal segment length $L_{\mathrm{opt}}^\varepsilon$ as well as the observed threshold. As an example, the case of $\eta_\text{cpl}=0.96$ is included in Fig.~\ref{fig:Lopt_vs_squeezing}.


\subsubsection{Achievable storage time}

\begin{figure}[t]
    \centering
    \begin{subfigure}[b]{1\linewidth}
        \centering
        \includegraphics[width=1\textwidth]{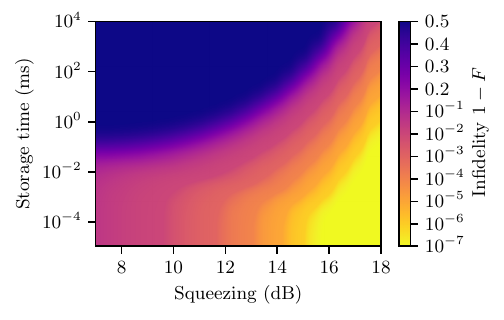}
        \vspace*{-59mm}
        \caption{\hspace*{-2.7mm}\phantom{a})}
        \vspace*{48mm}
        \label{fig:storage_fidelity}
    \end{subfigure}
    

    \begin{subfigure}[b]{1\linewidth}
        \centering
        \includegraphics[width=\textwidth]{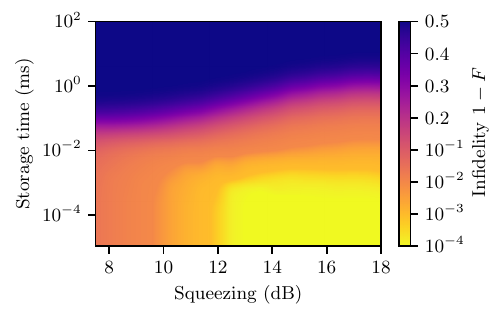}
        \vspace*{-59mm}
        \caption{\hspace*{-2.9mm}\phantom{b})}
        \vspace*{48mm}
        \label{fig:storage_fidelitycoupling loss}
    \end{subfigure}

    \caption{Optimized memory performance with a) no coupling loss and b) a coupling loss of $\eta_\text{cpl}=0.96$. The average infidelity $1-F_\text{avg}$ of the proposed quantum memory is plotted for different storage times $t_s$ and GKP squeezing $\varepsilon$. The infidelities are calculated using the optimal segment lengths $L^\varepsilon_{\mathrm{opt}}$ as well as the optimal decoder $D^{\varepsilon,\eta}_\text{opt}$ with $\eta=\eta\left(L^\varepsilon_{\mathrm{opt}}\right)$. Corrective Pauli operations are tracked and only applied at the end.}
    \label{fig:global_figure2}
\end{figure}

The optimized memory performance as a function of storage time and GKP squeezing is shown in Fig.~\ref{fig:storage_fidelity}.
For low squeezing levels, even millisecond storage times lead to fidelities approaching $0.5$, corresponding to the complete loss of logical information. In contrast, for sufficiently high squeezing the protocol supports significantly longer storage times, extending up to the second regime. We find that for a fixed target fidelity, the achievable storage time increases approximately exponentially with the available GKP squeezing. The strong dependence of the quantum memory performance on squeezing highlights the central importance of high-quality GKP resource states for the performance of optical quantum memories based on this architecture. Moreover, even in the limit of negligible storage times (corresponding to $n=1$), the single teleportation-based error correction step introduces a non-negligible fidelity penalty when the squeezing is low. This reflects the intrinsic noise associated with finite-energy GKP states and teleportation-based correction. Coupling losses severely affect the achievable storage time and performance of the memory as showcased in Fig.~\ref{fig:storage_fidelitycoupling loss} for a coupling loss of $\eta_\text{cpl}=0.96$. 

An additional route to extending storage times without requiring higher squeezing values is to concatenate the GKP code with a higher-level topological error correction code. Optical architectures enabling such concatenation have recently been proposed in \cite{aghaee_rad_scaling_2025, ostergaard_octo-rail_2025, haussler_long-distance_2025}. In these schemes, the discrete logical errors introduced by the GKP layer are corrected by the outer topological code. The currently known squeezing threshold above which such concatenated architectures become effective lies at $9.75\,\mathrm{dB}$, suggesting that moderate improvements in experimentally achievable GKP squeezing could enable dramatically longer storage times in future implementations.

\section{Discussion and Conclusion}\label{sec4}
We have presented a comprehensive quantitative analysis of a fully optical quantum-memory architecture in which GKP-encoded qubits are stored in a fibre loop and periodically stabilized using teleportation-based GKP error correction with deferred Pauli corrections. Propagation was modelled as a pure-loss channel which in combination with the following GKP error correction was represented as a syndrome-conditioned logical map acting on the Bloch vector of the encoded qubit. This description allowed us to construct a fidelity-optimized decoder and, crucially, to evaluate the full multi-round memory channel by explicitly composing the Pauli-frame sectors across successive rounds.

At the single-round level, the performance of the protocol reflects a fundamental competition between two noise mechanisms: loss-induced displacement noise accumulated during propagation and intrinsic finite-squeezing noise injected by teleportation-based correction. The optimized decoder reshapes the logical decision regions in phase space in a squeezing- and loss-dependent manner, yielding systematically improved average channel fidelities compared to a fixed nearest-lattice rule. This demonstrates that decoding is not merely a classical post-processing step but an integral part of the physical memory design, particularly in the experimentally relevant finite-squeezing regime.

For multi-round storage, it is shown that performance depends crucially on the segment length of the fibre delay line. Remarkably, this optimal segment length is approximately independent of the total storage length, providing a simple and robust design principle: for a given squeezing resource, one can choose a correction spacing that remains near-optimal across a broad range of storage times. Physically, this reflects a balance between suppressing accumulated loss noise and avoiding excessive injection of teleportation noise. A key qualitative feature of this analysis is the presence of a squeezing threshold around $6.7\,\mathrm{dB}$, which marks the transition between a regime where intermediate correction is beneficial and one where it becomes counterproductive. Below this value, the optimal segment length diverges and intermediate correction degrades performance. We also note that introducing additional noise due to coupling shifts this value towards higher squeezing.

The optimized storage map further shows that, for a fixed target fidelity, the achievable storage time increases approximately exponentially with squeezing. At low squeezing, even short storage times rapidly approach the $F_{\mathrm{avg}} \approx 1/2$ limit associated with complete loss of logical information. In contrast, higher squeezing enables substantially longer storage times when correction is applied at the appropriate spacing. These results translate improvements in squeezing directly into memory-lifetime gains and provide quantitative engineering targets for near-term photonic implementations. Similar trends and effects remain valid while dealing with coupling loss but the overall performance is heavily affected.

The present analysis assumes ideal internal components and models propagation as pure loss. In the future, the modelling of additional imperfections will be important to establish complete hardware-level error budgets. Such refinements can easily be incorporated within the presented framework by modifying the syndrome-conditioned logical map and re-optimizing the decoder accordingly.

Finally, there is a clear route toward extending storage times beyond the regime accessible with a single bosonic code. Once the squeezing exceeds the relevant threshold for code concatenation, the GKP encoding can be combined with a higher-level qubit error-correcting code to suppress residual logical noise. Together with continued improvements in deterministic preparation of high-quality finite-energy GKP states and advances in integrated photonic platforms, this approach offers a scalable path toward fault-tolerant, all-optical quantum memory compatible with deployed fibre infrastructure.

\begin{acknowledgments}
    We acknowledge support from the Danish National Research Foundation (bigQ, DNRF0142), Innovation Fund Denmark (QuantERA - ClusSTAR, 3155-00024A), EU ERC Adv (ClusterQ  grant no. 101055224) and EU Horizon Europe (CLUSTEC, grant agreement no. 101080173)
\end{acknowledgments}

\onecolumngrid
\appendix

\section{GKP teleportation as linear transformation of the Bloch vector}\label{sec:app}
The position-space wave function of the approximate GKP qunaught state is given by
\begin{align}
    \prescript{}{q}{\langle x|}\varnothing
    ^\varepsilon\rangle&= N_{\varnothing}\cdot\sum_{s\in\mathbb{Z}}\exp\left({-\Delta^2\pi s^2-\frac{(x-\lambda\sqrt{2\pi}s)^2}{2\Delta^2}}\right)
\end{align}
with $\Delta^2=\tanh\beta$ and $\lambda=(\cosh\beta)^{-1}$ \cite{matsuura_equivalence_2020}. Mixing two on a balanced beam-splitter then produces an approximate GKP Bell pair
\begin{align}\begin{split}
    \prescript{}{q_1,q_2}{\langle y, z|}\hat B_{\frac{\pi}{4}}\ket{\varnothing^\varepsilon\varnothing^\varepsilon}&=\prescript{}{q_1,q_2}{\left\langle \frac{y-z}{\sqrt{2}}, \frac{y+z}{\sqrt{2}}\bigg|\varnothing^\varepsilon\varnothing^\varepsilon\right\rangle}\\
    &={N}_\varnothing^2\sum_{s,t\in \mathbb{Z}}
     \exp\left({-\Delta^2\pi(s^2+t^2)}{-\frac{1}{2\Delta^2}\left(y-\lambda\sqrt{\pi}(s+t)\right)^2-\frac{1}{2\Delta^2}\left(z-\lambda\sqrt{\pi}(s-t)\right)^2}\right)\\
    &={N}_\varnothing^2\left(\sum_{u, v\text{ even}} \exp\left({-\frac{\Delta^2}{2}\pi u^2-\frac{1}{2\Delta^2}(y-\lambda\sqrt{\pi}u)^2}{-\frac{\Delta^2}{2}\pi v^2-\frac{1}{2\Delta^2}(z-\lambda\sqrt{\pi}v)^2}\right)\right.\\
    &\left.+\sum_{u,v\text{ odd}} \exp\left({-\frac{\Delta^2}{2}\pi u^2-\frac{1}{2\Delta^2}(y-\lambda\sqrt{\pi}u)^2}{-\frac{\Delta^2}{2}\pi v^2-\frac{1}{2\Delta^2}(z-\lambda\sqrt{\pi}v)^2}\right)\right)\\
    &=\frac{{N}_\varnothing^2}{{N}_{\bar 0}^2}\prescript{}{q_1,q_2}{\langle y, z|\bar 0^\varepsilon\bar 0^\varepsilon\rangle}+\frac{{N}_\varnothing^2}{{N}_{\bar 1}^2}\prescript{}{q_1,q_2}{\langle y, z|\bar 1^\varepsilon\bar 1^\varepsilon\rangle}
\end{split}\end{align}
with the appropriate norms $N_{\bar 0}$ and $N_{\bar 1}$.
Consequently, the output state $\ket{\psi_\text{out}(q_0, p_0)}$ of the teleportation circuit of Eq.~\eqref{Cir:teleport} for an arbitrary input state $\ket{\psi_\text{in}}$ is given by
\begin{align}\begin{split}
    \left\langle x |\psi_\text{out}(q_0,p_0) \right\rangle &= \frac{1}{\sqrt{2\pi P(q_0,p_0)}}\int dy\ e^{-ip_0y}  \prescript{}{q}{\left\langle \frac{y+q_0}{\sqrt{2}}\bigg|\psi_\text{in}\right\rangle}\prescript{}{q_1,q_2}{\left\langle \frac{y-q_0}{\sqrt{2}}, x\right|}\hat B_{\frac{\pi}{4}}\bigg|\varnothing^\varepsilon\varnothing^\varepsilon\bigg\rangle\\
    &= \frac{1}{\sqrt{\pi P(q_0,p_0)}}\int dy'\  e^{-i\sqrt{2}p_0y'} e^{-ip_0q_0} \prescript{}{q}{\left\langle y'+\sqrt{2}q_0\Big|\psi_\text{in}\right\rangle}\prescript{}{q_1,q_2}{\langle y', x|}\hat B_{\frac{\pi}{4}}\ket{\varnothing^\varepsilon\varnothing^\varepsilon}\\
    &=  \frac{1}{\sqrt{\pi P(q_0,p_0)}}\Bigg(\frac{N_\varnothing^2}{{N}_{\bar 0}^2}\bra{\bar{0}^\varepsilon}\hat{D}(-q_0-ip_0)\ket{\psi_\text{in}}\langle x \ket{\bar{0}^\varepsilon} +\frac{N_\varnothing^2}{{N}_{\bar 1}^2}\bra{\bar{1}^\varepsilon}\hat{D}(-q_0-ip_0)\ket{\psi_\text{in}}\langle x \ket{\bar{1}^\varepsilon}
    \Bigg)\\
    &=\frac{1}{\sqrt{\pi P(q_0,p_0)}}\langle x |\hat \Pi_\varnothing^\varepsilon\hat{D}(-q_0-ip_0)\ket{\psi_\text{in}}
\end{split}\end{align}
with the displacement operator $\hat{D}(q_0+ip_0)= e^{i\sqrt{2}p_0\hat{x}-i\sqrt{2}q_0\hat{p}}$ and the approximate GKP projector
\begin{align}\begin{split}
    \hat \Pi_\varnothing^\varepsilon\equiv
    \frac{{N}_\varnothing^2}{{N}_{\bar 0}^2}\ket{\bar 0^\varepsilon}\bra{\bar 0^\varepsilon}+\frac{{N}_\varnothing^2}{{N}_{\bar 1}^2}\ket{\bar 1^\varepsilon}\bra{\bar 1^\varepsilon}
    &=\left(\frac{{N}_\varnothing^2}{2{N}_{\bar 0}^2}+\frac{{N}_\varnothing^2}{2{N}_{\bar 1}^2}\right)\Big(\left(1+\delta\right)\ket{+^\varepsilon}\bra{+^\varepsilon}+\left(1-\delta\right)\ket{-^\varepsilon}\bra{-^\varepsilon}\Big)\\
    &+\sqrt{1-\delta^2}\left(\frac{{N}_\varnothing^2}{2{N}_{\bar 0}^2}-\frac{{N}_\varnothing^2}{2{N}_{\bar 1}^2}\right)\Big(\ket{+^\varepsilon}\bra{-^\varepsilon}+\ket{-^\varepsilon}\bra{+^\varepsilon}\Big).
\end{split}\end{align}
Hence, we can write the output state in the logical qubit basis
\begin{align}\begin{split}
    \ket{\psi_\text{out}(q_0,p_0)}
    &=\frac{1}{\sqrt{P(q_0,p_0)}}\left(c_+(q_0,p_0)\ket{+^\varepsilon}+c_-(q_0,p_0)\ket{-^\varepsilon}\right)
\end{split}\end{align}
using the coefficients
\begin{align}\begin{split}
    c_+(q_0,p_0)&
    =\frac{1}{\sqrt{\pi}}\bra{+^\varepsilon}\hat \Pi_\varnothing^\varepsilon\hat{D}(-q_0-ip_0)\ket{\psi_\text{in}},\\
    c_-(q_0,p_0)&
    =\frac{1}{\sqrt{\pi}}\bra{-^\varepsilon}\hat \Pi_\varnothing^\varepsilon\hat{D}(-q_0-ip_0)\ket{\psi_\text{in}}
\end{split}\end{align}
and the probability $P(q_0,p_0)=|c_+(q_0,p_0)|^2+|c_-(q_0,p_0)|^2$.
In case the input state $\ket{\psi_\text{in}}$ is also given in terms of basis coefficients, the GKP teleportation becomes a transformation of the logical qubit space which can be represented by the 4x4 matrix $T^{\varepsilon}(q_0,p_0)$ with entries
\begin{align}
    \left[T^{\varepsilon}(q_0,p_0)\right]_{ij}=\frac{1}{\pi}\Tr\left(\hat \sigma_i^\varepsilon\hat \Pi_\varnothing^\varepsilon\hat{D}(-q_0-ip_0)\hat \sigma^\varepsilon_j\hat{D}^\dagger(-q_0-ip_0)\hat \Pi_\varnothing^\varepsilon\right)
\end{align}
and its effect on the Bloch vector
\begin{align}
    \vec{a}_\text{out}(q_0,p_0)=T^{\varepsilon}(q_0,p_0) \vec{a}_\text{in}.
\end{align}
Note that, while $T^{\varepsilon}(q_0,p_0)$ is trace-preserving when integrating over $q_0$ and $p_0$ as
\begin{align}
    {a}_{\text{out}, 0}(q_0,p_0)=\frac{1}{\pi}\Tr\left(\hat \Pi_\varnothing^\varepsilon\hat{D}(-q_0-ip_0)\hat \rho^\varepsilon_\text{in}\hat{D}^\dagger(-q_0-ip_0)\hat \Pi_\varnothing^\varepsilon\right)=P(q_0,p_0),
\end{align}
it is not necessarily completely positive \cite{Jafarzadeh_2025}.
Finally, for an input state affected by photon loss the transformation matrix is given by
\begin{align}
    \left[T^{\varepsilon,\eta}(q_0,p_0)\right]_{ij}=\frac{1}{\pi}\Tr\left(\hat \sigma_i^\varepsilon\hat \Pi_\varnothing^\varepsilon\hat{D}(-q_0-ip_0)\mathcal L_\eta(\hat \sigma^\varepsilon_j)\hat{D}^\dagger(-q_0-ip_0)\hat \Pi_\varnothing^\varepsilon\right).
\end{align}

\bibliography{apssamp}
\end{document}